\documentclass[epj]{svjour}
\usepackage{graphics}
\begin{document}
\title{Functional derivatives of T$_c$ for a two-band superconductor: application to MgB$_{2}$}
\author{Bo\v zidar Mitrovi\' c\thanks{e-mail: mitrovic@brocku.ca}
}                     
%
%
\institute{Physics Department, Brock University, St. Catharines, Ontario, Canada L2S~3A1}
\date{Received: date / Revised version: date}
%
\abstract{ We address the question of how phonons of various frequencies 
contribute to the superconducting transition temperature $T_c$ of a multi-band 
superconductor by considering the functional derivatives of $T_c$ with 
respect to various intraband and interband electron-phonon coupling functions. 
A general scheme for computing such functional derivatives is developed. The proofs
are given that the functional derivatives which are diagonal in band indices 
are linear in phonon 
energy $\Omega$ at small $\Omega$, while the 
functional derivatives which are off-diagonal in band indices diverge 
at $\Omega$ = 0 as 1/$\Omega$. The case of a two-band model for MgB$_{2}$ is 
treated numerically. 
} 
\maketitle
\section{Introduction}
\label{intro}
Bergmann and Rainer [1] introduced an important diagnostic tool into 
the Eliashberg theory of superconductivity (for a pedagogical review of 
Eliashberg theory see [2]). They considered the functional derivative of 
the superconducting transition temperature $T_{c}$ with respect to the 
electron phonon coupling function $\alpha^{2}F(\Omega)$ of an isotropic 
(dirty) superconductor
\begin{eqnarray}
\frac{\delta T_{c}}{\delta\alpha^{2}F(\Omega)} & = & \lim_{\eta\rightarrow 0} 
\left(T_{c}[\alpha^{2}F(\Omega')+\eta\delta(\Omega'-\Omega)]\right. \nonumber\\
 &  & -\left.T_{c}[\alpha^{2}F(\Omega')]\right)/\eta\>.
\end{eqnarray}
The functional $T_{c}[\alpha^{2}F(\Omega)]$ is defined by the Eliashberg equations 
at $T_{c}$ [2]. The function $\delta T_{c}/\delta\alpha^{2}F(\Omega)$ 
provides answer to the question -- How are the phonons of frequency $\Omega$ 
effective in contributing to $T_{c}$? From the practical point of view the 
functional derivative $\delta T_{c}/\delta\alpha^{2}F(\Omega)$ gives the 
change $\Delta T_{c}$ in transition temperature when $\alpha^{2}F(\Omega)$ is 
changed by a {\em small} amount $\Delta\alpha^{2}F(\Omega)$ (say, by applying 
pressure [3], by alloying [4,5], or by implanting small concentrations of hydrogen 
into a metal [6])
\begin{equation}
\Delta T_{c} =\int_{0}^{+\infty}d\Omega\frac{\delta T_{c}}
{\delta\alpha^{2}F(\Omega)}\Delta \alpha^{2}F(\Omega)\>.
\end{equation}
The concept of functional derivative $\delta/\delta\alpha^{2}F(\Omega)$
was extended by Rainer and Bergmann [4] and others to several 
thermodynamic properties of isotropic superconductors
(for a review and an extensive list of references see [7]).

The main conclusions of Bergmann and Rainer [1] were that 
$\delta T_{c}/\delta\alpha^{2}F(\Omega)$ is always positive 
(they were able to prove this mathematically for the case 
when the Coulomb repulsion parameter $\mu^{*}$ [2] is zero), 
proportional to $\Omega$ at $\Omega\ll$2$\pi T_{c}$ (we choose units 
such that $\hbar$ = 1 and $k_{B}$ = 1) and with a maximum 
at $\Omega$ just above 2$\pi T_{c}$. Hence, the electron 
coupling to a phonon of any frequency has a positive contribution 
to $T_{c}$, but the small values of $\delta T_{c}/\delta\alpha^{2}F(\Omega)$
in the low frequency region imply that the changes of 
$\alpha^{2}F(\Omega)$ in this frequency range have no appreciable 
effect on $T_{c}$, in contrast to the influence of the low 
frequency part of $\alpha^{2}F(\Omega)$ on the electron-phonon 
coupling parameter $\lambda$ 
\begin{equation}
\lambda=2\int_{0}^{+\infty}d\Omega\alpha^{2}F(\Omega)/\Omega\>,
\end{equation}
which is used in McMillan-type interpolation formulae for $T_{c}$ [2].

Daams and Carbotte [8] considered the functional derivative of $T_{c}$ 
with respect to the Fermi surface averaged electron-phonon coupling 
function of an anisotropic superconductor and found that 
$\delta T_{c}/\delta\alpha^{2}F(\Omega)$ {\em diverges} at $\Omega$ = 0 as 
1$/\Omega$. Their explicit calculations for a separable model 
of anisotropy 
$\alpha^{2}F_{{\bf k},{\bf k}'}(\Omega)=($1$+a_{\bf k})
\alpha^{2}F(\Omega)($1$+a_{{\bf k}'})$, with the Fermi surface averages 
$\langle a_{\bf k}\rangle$ = 0 and $\langle a_{\bf k}^{2}\rangle\ll$1, 
showed that at small $\Omega$ $\delta T_{c}/\delta\alpha^{2}F(\Omega)$
goes {\em negative} and diverges as $-$1$/\Omega$. Thus in high 
purity anisotropic superconductors the electron coupling to low 
frequency phonons {\em decreases} $T_{c}$, which is analogous to 
the effect of elastic impurity scattering on transition temperature 
of anisotropic superconductors. However, Daams and Carbotte pointed out that 
for their choice of anisotropy parameter for Pb ($\langle a_{\bf k}^{2}\rangle$
 = 0.04) $\delta T_{c}/\delta\alpha^{2}F(\Omega)$ becomes negative only 
below 0.6$T_{c}$ where $\alpha^{2}F(\Omega)$ has very little weight 
and varies as $\Omega^{2}$, rendering 
1$/\Omega$-divergence in functional derivative harmless. The net 
effect of anisotropy in the pairing interaction is to increase the $T_{c}$
as the virtual scattering $({\bf k}\uparrow,-{\bf k}\downarrow)
\rightarrow({\bf k}'\uparrow,-{\bf k}'\downarrow)\rightarrow\cdots$ 
over the Fermi surface 
takes advantage of the regions where the pairing interaction is large. 

A consensus has emerged (for a review see [9]) that in order to 
describe the superconducting properties of a 40K superconductor 
MgB$_{2}$ the Eliashberg theory has to be applied to a multi-band 
case with [10] or without [11] gap anisotropy on different 
sheets of the Fermi surface. In [11] the Eliashberg equations for 
an effective two-band model of electronic structure and electron-phonon 
coupling in MgB$_{2}$ were solved and the calculated specific 
heat difference between the superconducting and the normal state was in 
good agreement with experiments over a wide temperature range below 
$T_{c}$. As a model for gap anisotropy the two-band model is the opposite 
extreme to the separable 
anisotropy considered in [8] -- there are four Eliashberg functions 
$\alpha^{2}F_{\sigma\sigma}(\Omega)$, $\alpha^{2}F_{\pi\pi}(\Omega)$, 
$\alpha^{2}F_{\sigma\pi}(\Omega)$ and $\alpha^{2}F_{\pi\sigma}(\Omega)$, 
and, correspondingly, there are four functional derivatives of $T_{c}$ with 
respect to each one of them (the band off-diagonal functions 
$\alpha^{2}F_{\sigma\pi}$ and $\alpha^{2}F_{\pi\sigma}$ are related, but 
are different as they are proportional to the partial electronic 
densities of states in $\pi$- and $\sigma$- bands, respectively). In this 
work we calculate the functional derivatives of $T_{c}$ for the two-band model  
and electron-phonon coupling functions  presented in [11].

The rest of the paper is organized as follows. In Section 2 we present the 
formalism necessary for computation of the functional derivatives of $T_{c}$ 
with respect to various electron-phonon coupling functions in a multi-band 
case. We also prove that the band-diagonal functional derivatives 
$\delta T_{c}/\delta\alpha^{2}F_{ii}(\Omega)$, where $i$ is the 
band index, are proportional to $\Omega$ at small $\Omega$, while the 
functional derivatives which are off-diagonal in the band indices,  
$\delta T_{c}/\delta\alpha^{2}F_{ij}(\Omega)$ with $i\neq j$, diverge 
at small $\Omega$ as 1$/\Omega$. In Section 3 we present and discuss
our numerical results, and in Section 4 we give a summary. 
\section{The functional derivatives $\delta T_{c}/\delta\alpha^{2}F_{ij}(\Omega)$ }
\label{sec:1}
In the case of several bands $i=1,2,\dots$ with different partial densities of states
$N_{i}$ the Eliashberg equations for $T_{c}$ do not have the form of a Hermitian 
eigenvalue problem [11]. That is because the interband electron-phonon coupling 
functions $\alpha^{2}F_{ij}(\Omega)$ and the corresponding Coulomb repulsion 
parameters $\mu_{ij}$ are proportional to $N_{j}$ and are not symmetric under the 
exchange of the band indices $i$ and $j$, $i\neq j$. As a calculation of the 
functional derivative of $T_{c}$ relies on the Hellman-Feynman theorem [1], [12]
which is valid only for Hermitian matrices, it is necessary to cast the 
Eliashberg equations at $T_{c}$ into a Hermitian eigenvalue problem. To this end 
one first takes the cutoff $\omega_{c}$ in the Matsubara frequency sums to be 
large enough so that $\mu_{ij}^{*}(\omega_{c})=\mu_{ij}\equiv V_{ij}^{c} N_{j}$, 
where $V_{ij}^{c}$ is the Fermi surface averaged screened Coulomb matrix 
element between the states in the bands $i$ and $j$; clearly $V_{ji}^{c}=V_{ij}^{c}$. 
The electron-phonon coupling functions can be written as $\alpha^{2}F_{ij}(\Omega)=
\alpha^{2}f_{ij}(\Omega)N_{j}$ with $\alpha^{2}f_{ij}(\Omega)=\alpha^{2}f_{ji}(\Omega)$. 
Then the Eliashberg equations at $T_{c}$ take the form 
\begin{equation}
\phi_{i}(n)  =  \pi T_{c}\sum_{jm}[{\bar \lambda}_{ij}(n-m)N_{j}-V_{ij}^{c} N_{j}]
\frac{\phi_{j}(m)}{|\omega_{m}|Z_{j}(m)}\>, 
\end{equation}
\begin{equation}
\omega_{n}Z_{i}(n)  =  \omega_{n}+\pi T_{c}\sum_{jm}{\bar \lambda}_{ij}(n-m)N_{j}
\frac{\omega_{m}}{|\omega_{m}|}\>, 
\end{equation}
\begin{equation}
{\bar \lambda}_{ij}(n-m)  =  \int_{0}^{+\infty}d\Omega 
\alpha^{2}f_{ij}(\Omega)\frac{2\Omega}{\Omega^{2}+
(\omega_{n}-\omega_{m})^{2}}\>,
\end{equation}
where $\phi_{i}(n)$ and $Z_{i}(n)$ are the pairing self-energy and the 
renormalization function, respectively, at Matsubara frequency $\omega_{n}=
\pi T_{c}(2n-1)$ in band $i$. By defining
\begin{equation}
{\bar \phi}_{i}(n)=\phi_{i}(n)\sqrt{N_{i}}/\sqrt{|\omega_{n}|Z_{i}(n)}
\end{equation}
Eq.~(4) takes the form of a 
Hermitian eigenvalue problem
\begin{equation}
{\bar \phi}_{i}(n)=\varepsilon(T)\sum_{jm}\pi T\frac{\lambda_{ij}^{s}(n-m)-\mu_{ij}^{s}}
{\sqrt{|\omega_{n}|Z_{i}(n)}\sqrt{|\omega_{m}|Z_{j}(m)}}{\bar \phi}_{j}(m)\>,
\end{equation}
where the symmetrized $\lambda$'s and Coulomb repulsion parameters are given by
\begin{equation}
\lambda_{ij}^{s}(n-m)=\sqrt{\frac{N_{i}}{N_{j}}}\int_{0}^{+\infty}\alpha^{2}F_{ij}(\Omega)
\frac{2\Omega}{\Omega^{2}+
(\omega_{n}-\omega_{m})^{2}}\>,
\end{equation}
\begin{equation}
\mu_{ij}^{s}=\sqrt{\frac{N_{i}}{N_{j}}}\mu_{ij}\>,
\end{equation}
and the eigenvalue $\varepsilon(T)$ is 1 when $T=T_{c}$. In terms of 
$\lambda_{ij}^{s}(n-m)$ the renormalization function $Z_{i}(n)$ is   
given by (see Eq.~(5))
\begin{equation}
\omega_{n}Z_{i}(n)  =  \omega_{n}+\pi T_{c}\sum_{jm}\lambda_{ij}^{s}(n-m)
\sqrt{\frac{N_{j}}{N_{i}}}\frac{\omega_{m}}{|\omega_{m}|}\>.
\end{equation}
Next, in order not
to deal with a matrix of unnecessarily large size one cuts off the Mutsubara 
sums in (8) at a smaller energy $\omega_{c}$, which is large enough so that 
$Z_{i}(n)\approx$1 for $|\omega_{n}|>\omega_{c}$, and at the same time 
rescales $\mu_{ij}^{s}$ to the new cutoff $\omega_{c}$ by integrating out 
the high energy part of ${\bar \phi}_{j}(m)$ as described in [2]. 
The result is that $\mu_{ij}^{s}$ in Eq.~(8) is replaced by $\mu_{ij}^{*}
(\omega_{c})$ where the matrix (in band indices) $\hat{\mu}^{*}(\omega_{c})$ is 
related to matrix $\hat{\mu}^{s}$ by 
\begin{equation}
\hat{\mu}^{*}(\omega_{c})=\left(\hat{1}+\hat{\mu}^{s}\ln\frac{E_{F}}{\omega_{c}}
\right)^{-1}\hat{\mu}^{s}\>,
\end{equation}
with $E_{F}$ on the order of the total bandwidth. In the case of a two-band 
model, which we will examine numerically in the next section, the explicit 
relations between $\mu_{ij}^{*}(\omega_{c})$ and $\mu_{ij}^{s}$ are 
($\mu_{\pi\sigma}^{s}=\mu_{\sigma\pi}^{s}$)
\begin{eqnarray}
\mu_{\sigma\sigma}^{*}(\omega_{c}) & = &  \left[\mu_{\sigma\sigma}^{s}+ 
(\mu_{\sigma\sigma}^{s}\mu_{\pi\pi}^{s}-{\mu_{\sigma\pi}^{s}}^{2})
\ln\frac{E_{F}}{\omega_{c}}\right]/D\>, \\
\mu_{\sigma\pi}^{*}(\omega_{c}) & = & \mu_{\sigma\pi}^{s}/D\>, \\
\mu_{\pi\pi}^{*}(\omega_{c}) & = &  \left[\mu_{\pi\pi}^{s}+
(\mu_{\sigma\sigma}^{s}\mu_{\pi\pi}^{s}-{\mu_{\sigma\pi}^{s}}^{2})
\ln\frac{E_{F}}{\omega_{c}}\right]/D\>,
\end{eqnarray}
where $D$ is the determinant of $\hat{1}+\hat{\mu}^{s}\ln(E_{F}/\omega_{c})$
\begin{eqnarray}
D & = & 1+(\mu_{\sigma\sigma}^{s}+\mu_{\pi\pi}^{s})\ln\frac{E_{F}}{\omega_{c}}
\nonumber \\
  &   &\times (\mu_{\sigma\sigma}^{s}\mu_{\pi\pi}^{s}-{\mu_{\sigma\pi}^{s}}^{2})
\left(\ln\frac{E_{F}}{\omega_{c}}\right)^{2}\>.
\end{eqnarray}

Now it is easy to generalize the procedure for calculating the functional 
derivative of $T_{c}$ given in [12] for a single-band isotropic superconductor 
to the case of several isotropic bands. One finds 
\begin{equation}
\frac{\delta T_{c}}{\delta\alpha^{2}F_{ij}(\Omega)}=
-\left(\frac{d\varepsilon(T_{c})}{dT}\right)^{-1}
\frac{\delta \varepsilon(T_{c})}{\delta\alpha^{2}F_{ij}(\Omega)}
\end{equation}
with
\begin{eqnarray}
\frac{\delta\varepsilon(T_{c})}{\delta\alpha^{2}F_{ij}(\Omega)} & = &
\left[\pi T_{c}\sum_{j}\sum_{n,m=1}^{N_{c}}
\frac{\bar{\phi}_{i}(n)}{\sqrt{\omega_{n}Z_{i}(n)}}
\frac{\bar{\phi}_{j}(m)}{\sqrt{\omega_{m}Z_{j}(m)}}\right.\nonumber \\
&   & \times\sqrt{\frac{N_{i}}{N_{j}}}
\left(\frac{2\Omega}{\Omega^{2}+(2\pi T(n-m))^{2}}\right. \nonumber \\
&   & \left. +\frac{2\Omega}{\Omega^{2}+(2\pi T(n+m-1))^{2}}\right)-\pi T \nonumber \\
&   & \times \sum_{n=1}^{N_{c}}\frac{\bar{\phi}_{i}^{2}(n)}
{\omega_{n}Z_{i}(n)}\left(2\sum_{l=0}^{n-1}
\frac{2\Omega}{\Omega^{2}+(2\pi Tl)^{2}}\right. \nonumber \\
&   &\left.\left.-\frac{2}{\Omega}\right)\right]
/\sum_{k}\sum_{n=1}^{N_{c}}\bar{\phi}_{k}^{2}(n)\>.
\end{eqnarray}
The derivative $d\varepsilon(T_{c})/dT$($<$0) is conveniently calculated in the 
process of finding the highest T for which the largest $\varepsilon(T)$ in Eq.~(8) is 
equal to 1 and $\bar{\phi}_{i}(n)$ are the components of the corresponding 
eigenvector of length $dN_{c}$, where $d$ is the number of bands and 
$N_{c}=[\omega_{c}/(2\pi T_{c})+$0.5$]$, where 
$[\cdots]$ denotes the integer part. Note that $\bar{\phi}_{i}(n)/
\sqrt{\omega_{n}Z_{i}(n)}=\sqrt{N_{i}}\phi_{i}(n)/\omega_{n}Z_{i}(n)=
\sqrt{N_{i}}\Delta_{i}(n)/\omega_{n}$, where $\Delta_{i}(n)$ is the gap function 
in band $i$ at Matsubara frequency $i\omega_{n}$. 

For $\Omega\ll 2\pi T_{c}$ one finds from Eq.~(18) that the band-diagonal 
functional derivatives are given by 
\begin{eqnarray}
\frac{\delta\varepsilon(T_{c})}{\delta\alpha^{2}F_{ii}(\Omega)} & = & 2\pi T_{c}\Omega 
\left[\sum_{i}\sum_{n,m=1}^{N_{c}}
\frac{\bar{\phi}_{i}(n)}{\sqrt{\omega_{n}Z_{i}(n)}}
\frac{\bar{\phi}_{i}(m)}{\sqrt{\omega_{m}Z_{i}(m)}}\right.\nonumber \\
&   & \times\left(\frac{1}{(2\pi T_{c}(n-m))^{2}}+\right. \nonumber \\
&   &\left.\frac{1}{(2\pi T_{c}(n+m-1))^{2}}\right)\nonumber \\
&   &\left. - \sum_{n=2}^{N_{c}}\frac{\bar{\phi}_{i}^{2}(n)}
{\omega_{n}Z_{i}(n)}\sum_{l=1}^{n-1}\frac{1}{(2\pi T_{c}l)^{2}}\right] \nonumber \\
&   &/\sum_{k}\sum_{n=1}^{N_{c}}\bar{\phi}_{k}^{2}(n)\>,
\end{eqnarray}
and are {\em linear} in $\Omega$ just like in the one-band isotropic case [1]. The 
band-off-diagonal functional derivatives ($i\neq j$) in the small $\Omega$ limit
are given by
\begin{eqnarray}
\frac{\delta\varepsilon(T_{c})}{\delta\alpha^{2}F_{ij}(\Omega)}  & = & 
\frac{2\pi T_{c}}{\Omega}\sum_{n=1}^{N_{c}}
\frac{\bar{\phi}_{i}(n)}{\sqrt{\omega_{n}Z_{i}(n)}} \nonumber \\
& & \left(\sqrt{\frac{N_{i}}{N_{j}}}\frac{\bar{\phi}_{j}(n)}{\sqrt{\omega_{n}Z_{j}(n)}}-
\frac{\bar{\phi}_{i}(n)}{\sqrt{\omega_{n}Z_{i}(n)}}\right) \nonumber \\
&   &/\sum_{k}\sum_{n=1}^{N_{c}}\bar{\phi}_{k}^{2}(n)\>.
\end{eqnarray}
and diverge at $\Omega$ = 0 as 1/$\Omega$ just like the functional derivative 
of $T_{c}$ with respect to the Fermi surface averaged electron-phonon 
coupling function of an anisotropic superconductor [8]. Note that the sum in the 
numerator of Eq.~(20) can be written as $N_{i}\sum_{n=1}^{N_{c}}\Delta_{i}(n)
(\Delta_{j}(n)-\Delta_{i}(n))/\omega_{n}^{2}$ and the sign of 
$\delta T_{c}/\delta\alpha^{2}F_{ij}(\Omega)$ is determined by the relative size 
of the gaps $\Delta_{i}(n)$ and $\Delta_{j}(n)$ {\em near} $T_{c}$ 
in the two bands and by the sign of 
of $\Delta_{i}(n)$ {\em near} $T_{c}$ for low $n$ since the low-$n$ terms give the 
largest contribution to the sum because of $\omega_{n}^{2}$ in the denominator. 

The results presented so far apply to a superconductor with {\em any} number 
of bands with isotropic intraband and interband interactions (both electron-phonon 
and Coulomb). In the next section we present numerical results for an 
effective two-band model of MgB$_{2}$ described in [11]. 
\section{Numerical results for a two-band model}
\label{sec:2}
Following the work of Liu {\em at al} [13], Golubov {\em at al} [11] reduced the 
four-band electronic structure and electron-phonon coupling in MgB$_{2}$ to an 
effective two-band model by exploiting the similarity of the two cylindrical 
($\sigma$-bands) and the two three-dimensional ($\pi$-bands) sheets of the Fermi 
surface. We used the $\alpha^{2}F$'s given in [11] with the coupling parameters 
(see Eq.~(3)) $\lambda_{\sigma\sigma}$ = 1.017, $\lambda_{\pi\pi}$ = 0.446, 
$\lambda_{\sigma\pi}$ = 0.212 and $\lambda_{\pi\sigma}$ = 0.155. The Coulomb 
repulsion parameters were determined using the ratios of the {\em screened} 
Coulomb interaction parameters for MgB$_{2}$ calculated in [14], $\mu_{\sigma\sigma}$ : 
$\mu_{\pi\pi}$ : $\mu_{\sigma\pi}$ : $\mu_{\pi\sigma}$ = 1.75 : 2.04 : 1.61 : 1.00, 
the density of states ratio $N_{\pi}/N_{\sigma}$ consistent with the ratio 
$\lambda_{\sigma\pi}/\lambda_{\pi\sigma}$ for the spectra that we used, and using 
equations (10) and (13-16) with $E_{F}$ set equal to the $\pi$-bandwidth 
of 15 eV [15] and $\omega_{c}$ = 0.5 eV. These constraints leave the single 
fitting parameter $\mu_{\sigma\sigma}$ which was fitted to the experimental 
transition temperature of 39.4 K. The results of the fit were: $\mu_{\sigma\sigma}^{*}
(\omega_{c})$ = 0.19627, $\mu_{\pi\pi}^{*}(\omega_{c})$ = 0.19561 and 
$\mu_{\sigma\pi}^{*}(\omega_{c})$ = $\mu_{\pi\sigma}^{*}(\omega_{c})$ = 0.04948. 

The calculated functional derivatives 
are shown in Figures 1-3.    
%
\begin{figure}
\resizebox{0.45\textwidth}{!}{%
  \includegraphics{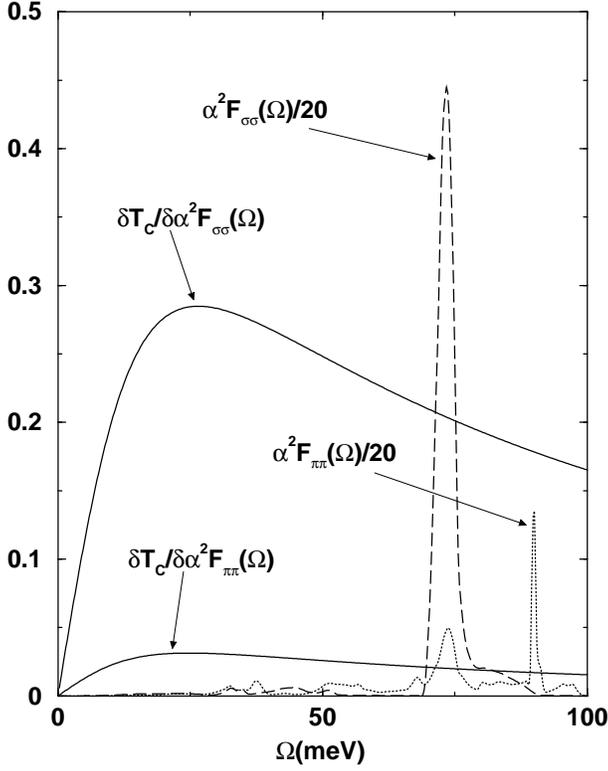}
  }
  \caption{The band-diagonal functional derivatives and rescaled electron-phonon coupling
  functions over the entire phonon energy range in MgB$_{2}$.}
  \label{fig:1}       
  \end{figure}
  \begin{figure}
  \resizebox{0.45\textwidth}{!}{%
  \includegraphics{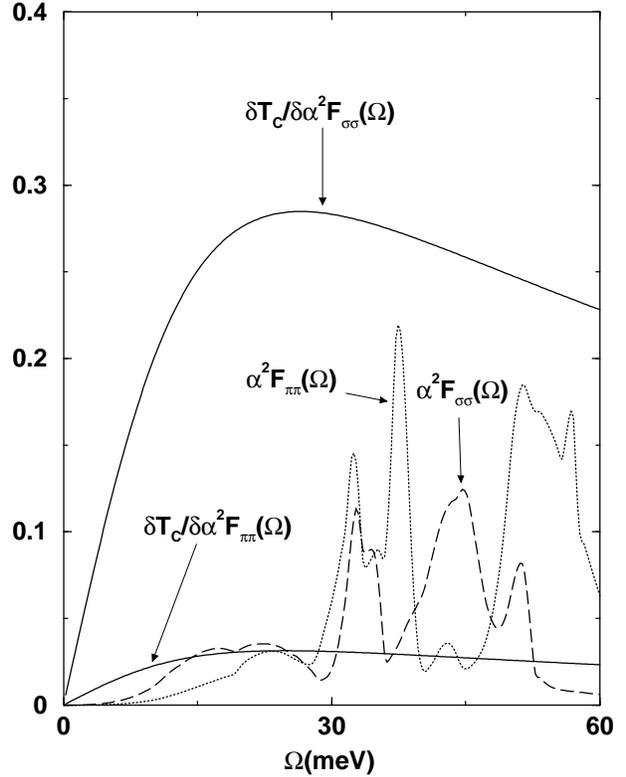}
  }
  \caption{The same as Fig.~1 but over a smaller energy range so that the the parts
  of $\alpha^{2}F_{\sigma\sigma}(\Omega)$ and $\alpha^{2}F_{\pi\pi}(\Omega)$ that
  are near the maxima in the corresponding functional derivatives are drawn to scale.}
  \label{fig:2}
\end{figure}
\begin{figure}
\resizebox{0.45\textwidth}{!}{%
\includegraphics{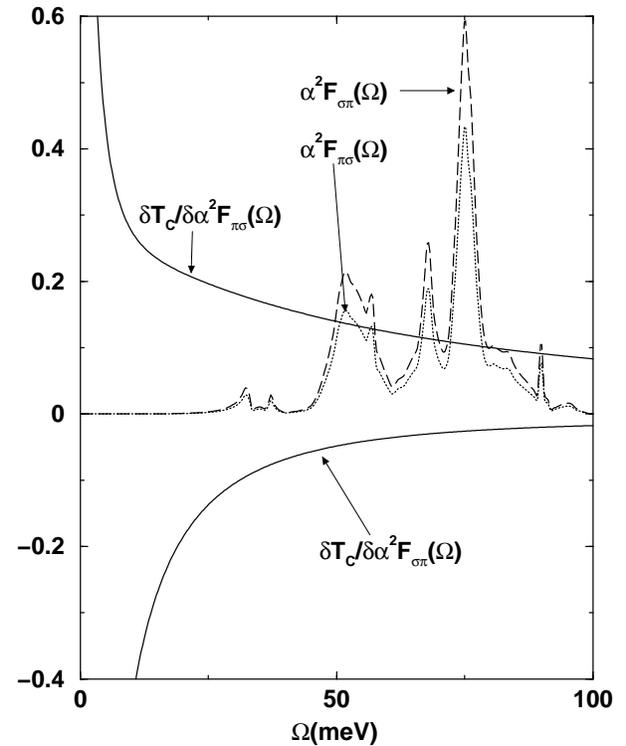}
}
\caption{The band-off-diagonal functional derivatives and electron-phonon coupling
functions for MgB$_{2}$.}
\label{fig:3}
\end{figure}
In Fig.~1 we present the band-diagonal functional derivatives 
$\delta T_{c}/\delta\alpha^{2}F_{\sigma\sigma}(\Omega)$, 
$\delta T_{c}/\delta\alpha^{2}F_{\pi\pi}(\Omega)$ and rescaled 
$\alpha^{2}F_{\sigma\sigma}(\Omega)$ and $\alpha^{2}F_{\pi\pi}(\Omega)$ over the 
entire phonon energy range in MgB$_{2}$. The spectra 
$\alpha^{2}F_{\sigma\sigma}(\Omega)$ and $\alpha^{2}F_{\pi\pi}(\Omega)$ are 
scaled down by a factor of 20 so that their shapes over the entire range 
of phonon energies are seen on the scale set by the size of the 
corresponding functional derivatives. This leaves parts of 
$\alpha^{2}F_{\sigma\sigma}(\Omega)$ and $\alpha^{2}F_{\pi\pi}(\Omega)$ 
that are near the maxima in corresponding functional derivatives 
invisible, and in Fig.~2 we redraw a part of Fig.~1 over a smaller 
energy range with the electron-phonon coupling functions shown to scale. 
As we proved in Sect.~\ref{sec:1},   
$\delta T_{c}/\delta\alpha^{2}F_{\sigma\sigma}(\Omega)$ 
and $\delta T_{c}/\delta\alpha^{2}F_{\pi\pi}(\Omega)$ are linear in $\Omega$ 
at small energies and vanish at $\Omega$ = 0. Moreover,  
both functional derivatives are positive with broad maxima at 9-10 times 
$k_{B}T_{c}$, similar to what one finds in 
the isotropic single-band case [1],[12]. The difference 
in sizes of the two functional derivatives in Figs.~1 and 2 is due to the difference 
in sizes of the gap-functions $\Delta_{\sigma}(n)$ and $\Delta_{\pi}(n)$ near $T_{c}$
in the two bands as can be deduced from Eq.~(18) (or Eq.~(19)) by noting 
that the normalization factor $\sum_{n}({\bar \phi}_{\sigma}^{2}(n)+
{\bar \phi}_{\pi}^{2}(n))$ is the same for both functional derivatives and it's 
size is largely determined by ${\bar \phi}_{\sigma}(n)$, which is larger than  
${\bar \phi}_{\pi}(n)$ (${\bar \phi}_{\sigma}($1$)/{\bar \phi}_{\pi}($1$)$ = 3.4).
Note that in the isotropic one-band case the
scale of the functional derivative of $T_{c}$ varies roughly as 1/(1+$\lambda$) 
[16] and from Fig.~1 it is clear that such a ``rule'' cannot be applied in 
determining the relative size of the band-diagonal functional derivatives of 
$T_{c}$ in a multi-band case.

In Fig.~3 we present the band-off-diagonal functional derivatives 
$\delta T_{c}/\delta\alpha^{2}F_{\sigma\pi}(\Omega)$ and 
$\delta T_{c}/\delta\alpha^{2}F_{\pi\sigma}(\Omega)$ together 
with the corresponding electron-phonon coupling functions. 
As we proved in Sect.~\ref{sec:1}, both of these functional 
derivatives diverge at $\Omega$ = 0 as 1/$\Omega$, but they also have 
opposite signs. Both $\alpha^{2}F_{\sigma\pi}(\Omega)$ and 
$\alpha^{2}F_{\pi\sigma}(\Omega)$ vary as $\Omega^{2}$ in the limit 
$\Omega\rightarrow$ 0 so that 1/$\Omega$-divergences in the corresponding 
functional derivatives are integrable, as can be deduced from Eq.~(2).

%
The difference in signs between $\delta T_{c}/\delta\alpha^{2}F_{\sigma\pi}(\Omega)$
and $\delta T_{c}/\delta\alpha^{2}F_{\pi\sigma}(\Omega)$ is related to the fact that
near $T_{c}$ 
$\Delta_{\sigma}(n)>\Delta_{\pi}(n)$ with both gap functions {\em positive}
at low $n$ (see Eq.~(20) and the subsequent discussion in Sect.~\ref{sec:1}). In
order to illustrate the importance of the sign of the smaller gap $\Delta_{\pi}(n)$
at low $n$ close to $T_{c}$,  
we have computed the functional derivatives for the case when
$\alpha^{2}F_{\pi\pi}(\Omega)$, $\alpha^{2}F_{\sigma\pi}(\Omega)$ and
$\alpha^{2}F_{\pi\sigma}(\Omega)$ were scaled down by a factor of 10, with
$\alpha^{2}F_{\sigma\sigma}(\Omega)$ and $\mu_{\sigma\sigma}^{*}(\omega_c)$,
$\mu_{\pi\pi}^{*}(\omega_c)$,
$\mu_{\sigma\pi}^{*}(\omega_c)$, $\mu_{\pi\sigma}^{*}(\omega_c)$ left unchanged.
This produced negative $\lambda_{\pi\pi}-\mu_{\pi\pi}^{*}$, 
$\lambda_{\pi\sigma}-\mu_{\pi\sigma}^{*}$ and 
$\lambda_{\sigma\pi}-\mu_{\sigma\pi}^{*}$ which
resulted in a solution where $\Delta_{\sigma}(n)$ and $\Delta_{\pi}(n)$ near $T_{c}$ 
have opposite signs at low $n$. The corresponding functional derivatives are shown in 
Figure 4. Now, both band-off-diagonal functional derivatives 
are negative. Note that the scale of $\delta T_{c}/\delta\alpha^{2}F_{\pi\pi}$ 
is roughly two orders of magnitude smaller than the other three functional 
derivatives.

It is important to point out that the calculated $T_{c}$ for the parameters
\begin{figure}
\resizebox{0.45\textwidth}{!}{%
\includegraphics{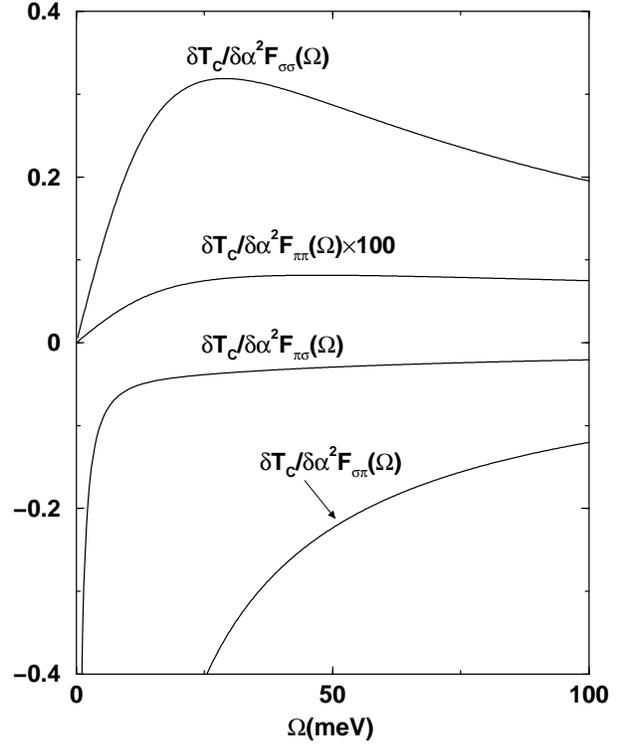}
}
\caption{The functional derivatives for the case when the strengths of
$\alpha^{2}F_{\pi\pi}$, $\alpha^{2}F_{\sigma\pi}$ and $\alpha^{2}F_{\pi\sigma}$
were scaled down by a factor of 10 with $\alpha^{2}F_{\sigma\sigma}$ and
$\mu^{*}(\omega_c)$'s left unchanged compared to those used in Figs.~1-3.}
\label{fig:4}
\end{figure}
in Fig.~4 was 43.7 K -- substantially higher (by 10\%) than the $T_{c}$ of 
39.4 K obtained for the parameters used to produce the results given in 
Figs.~1-3. In fact, even higher $T_{c}$ of 45.1 K was obtained by setting 
$\alpha^{2}F_{\pi\pi}$, $\alpha^{2}F_{\sigma\pi}$ and $\alpha^{2}F_{\pi\sigma}$ 
identically equal to 0 with $\alpha^{2}F_{\sigma\sigma}$ and all the Coulomb 
repulsion parameters left the same - i.e.~no attractive interaction in 
$\pi$-$\pi$ and $\sigma$-$\pi$ channels!
The corresponding functional 
derivatives were similar to those shown in Fig.~4 and they explain why $T_{c}$ is 
{\em reduced} as the couplings $\alpha^{2}F_{\pi\pi}$, $\alpha^{2}F_{\sigma\pi}$ 
and $\alpha^{2}F_{\sigma\pi}$ are turned on from 0: 
$\delta T_{c}/\delta\alpha^{2}F_{\sigma\pi}$ and 
$\delta T_{c}/\delta\alpha^{2}F_{\pi\sigma}$ are both negative and 
much bigger in absolute value than than the positive $\delta T_{c}/\delta\alpha^{2}F_{
\pi\pi}$. We finally note that if all interactions, both electron-phonon and Coulomb, in 
$\pi$-$\pi$ and $\sigma$-$\pi$ channels are set equal to zero, which effectively 
reduces the the two-band model to one-band model, the calculated $T_{c}$ was
44.6 K. This is a half degree {\em lower} than what was obtained by turning off only 
electron-phonon interaction in $\pi$-$\pi$ and $\sigma$-$\pi$ channels. 
\section{Summary}
\label{sec:3}
We have developed the general formalism for calculating the 
functional derivatives of $T_{c}$ with respect 
to electron-phonon coupling functions for a superconductor 
with several bands with isotropic intraband and interband 
interactions (electron-phonon and Coulomb). We proved rigorously 
that the band-diagonal functional derivatives 
$\delta T_{c}/$\newline$\delta\alpha^{2}F_{ii}(\Omega)$ are linear in 
$\Omega$ at small $\Omega$, as in the isotropic single band 
case [1]. At the same time we proved that the functional 
derivatives which are off-diagonal in the band indices, 
$\delta T_{c}/\delta\alpha^{2}F_{ij}(\Omega)$ with $i\neq j$, diverge 
at small $\Omega$ as 1/$\Omega$. The calculation was carried out  
for a two-band model of MgB$_2$ using the electron-phonon 
coupling spectra given in [11] and the ratios of the screened Coulomb 
interaction parameters given in [14].
We found that the band-diagonal functional derivatives are both 
positive with broad maxima in the range of 9-10 times $k_B T_c$, similar 
to the single band isotropic case. However, the functional derivative 
with respect to intraband electron-phonon coupling function for the 
band with the smaller gap ($\pi$ band) was found to be much smaller 
than the corresponding functional derivative for the band with the larger 
gap ($\sigma$ band). The functional derivatives with respect to the interband 
electron-phonon coupling functions were found to diverge as 1/$\Omega$ at 
$\Omega$ = 0, but had opposite signs over the entire range of phonon 
energies for the parameters given in [11]. We found that the signs of 
these off-diagonal functional derivatives are determined by the relative signs 
of the gap functions near $T_{c}$ at low Matsubara frequencies in the two bands and 
that, in general, it is possible to have both band-off-diagonal functional 
derivatives negative. 

The results found here give a better insight into the questions - What is the 
effect of phonons of frequency $\Omega$ on $T_{c}$ through their couplings  
to electrons via various band channels? Are all coupling always contributing 
positively to $T_{c}$, or are some of them pair-breaking? The answers to these 
questions are provided in Figures 1-3.\\

This work was supported by the Natural Sciences and
Engineering Research Council of Canada. We are grateful to O.~Jepsen for
providing the numerical values of $\alpha^{2}F$'s for MgB$_{2}$ presented 
in [11] and to S.~K.~Bose and K.~V.~Samokhin for their interest in this work.

\label{sec:2}
%
%
%
%
%

\end{document}